\documentclass[twocolumn,showpacs,showkeys,amsmath,amssymb,superscriptaddress,floatfix,nofootinbib]{revtex4}

\usepackage{graphicx}
\usepackage{bm} 
\usepackage{subfigure}
	
\def\vec#1{\mathchoice{\mbox{\boldmath$\displaystyle#1$}}
{\mbox{\boldmath$\textstyle#1$}}
{\mbox{\boldmath$\scriptstyle#1$}}
{\mbox{\boldmath$\scriptscriptstyle#1$}}}
\makeatletter

\newcommand{\pal}{$p$$\bar{\Lambda}$~}
\newcommand{\pla}{$p$$\Lambda$~}
\newcommand{\apla}{$\bar{p}$$\Lambda$~}
\newcommand{\apal}{$\bar{p}$$\bar{\Lambda}$~}
\newcommand{\bab}{$B\bar{B}$~}
\newcommand{\bb}{$BB$~}
\newcommand{\pap}{$p$$\bar{p}$~}

\begin{document}
 
\title{Extracting baryon-antibaryon strong interaction potentials from
  p$\bar{\Lambda}$ femtoscopic correlation function.%
\footnote{Supported by the Polish National Science Centre Grant
  No. 2011/01/B/ST2/03483 and 2012/07/D/ST2/02123}}

\author{Adam Kisiel} 
\email{kisiel@if.pw.edu.pl}
\affiliation{Faculty of Physics, Warsaw University of Technology, 
ul. Koszykowa 75, 00-662, Warsaw, Poland}

\author{Hanna Zbroszczyk} 
\affiliation{Faculty of Physics, Warsaw University of Technology, 
ul. Koszykowa 75, 00-662, Warsaw, Poland}

\author{Maciej Szyma\'nski} 
\affiliation{Faculty of Physics, Warsaw University of Technology, 
ul. Koszykowa 75, 00-662, Warsaw, Poland}

 
\begin{abstract}
The STAR experiment has measured $p\Lambda$, $\bar{p}\bar{\Lambda}$,
$\bar{p}\Lambda$, and $p\bar{\Lambda}$ 
femtoscopic  correlation functions in central Au+Au collisions at
$\sqrt{s_{NN}}=200$~GeV. The system size extracted for \pla and \apal
is consistent with model expectations and results for other pair
types, while for \pal and \apla it is not consistent with the
other two and significantly lower. In addition an attempt was made to
extract the unknown parameters of the strong 
interaction potential for this baryon-antibaryon ($B\bar{B}$) pair. In
this work we reanalyze the STAR data, taking into account residual
femtoscopic correlations from heavier $B\bar{B}$ pairs. We obtain new
estimates for the system size, consistent with the results for \pla and
\apal pairs and with model expectations. We give new estimates for the
strong interaction potential parameters for \pal and show that similar
constraints can be given for parameters for other, heavier
$B\bar{B}$ pairs. 
\end{abstract}



\maketitle 


\section{Introduction} 
\label{sec:intro} 
Strong interaction in a two-baryon system is one of the fundamental
problems in QCD~\cite{Rijken:2010zzb,Haidenbauer:2013aj}. Such
processes are measured in dedicated
experiments~\cite{Bruckner:1985zi,Bugg:1987nq,Bruckner:1991cq} and
significant 
body of data exists for baryon-baryon (BB)
interactions~\cite{Beringer:1900zz}. Baryon-antibaryon ($B\bar{B}$)
interaction includes a contribution from matter-antimatter
annihilation. This process for $p$$\bar{p}$ was studied in
great detail
theoretically~\cite{Batty:1989gg,Pirner:1991mn,Grach:1997mc,Klempt:2002ap}
and is measured with good
precision~\cite{Beringer:1900zz}. However no measurement exist for any
$B\bar{B}$ system other than 
$p$$\bar{p}$, $p$$\bar{n}$ and $\bar{p}$$d$. There is also 
little theoretical guidance on what to expect for \bab interaction for
other baryon types. The standard hadronic rescattering code used in
heavy-ion collision modeling, UrQMD~\cite{Bleicher:1999xi}, assumes
that any \bab interaction has the same parameters as the
$p$$\bar{p}$, expressed either as a function of relative momentum or
$\sqrt{s}$ of the pair. 

The STAR experiment has measured \pal femtoscopic
correlation~\cite{Adams:2005ws} in Au+Au collisions at
$\sqrt{s_{NN}}=200$~GeV. In that work a novel method was proposed to
determine the parameters of the strong interaction potential for \bab
pairs, using such correlations~\cite{Lednicky:1981su}. An estimate for
the real and imaginary part 
of the scattering length $f_{0}$ was given, showing significant
imaginary component, reflecting \bab annihilation in this 
channel. At the same time femtoscopic system size (radius) was
extracted. Surprisingly it was 50\% lower then the one for
regular $BB$ pairs at similar pair transverse mass $m_{T}$. It was
also inconsistent with hydrodynamic model predictions, which give
approximate scaling of the radii with $1/\sqrt{m_{T}}$. This scaling is
in agreement with all other femtoscopic measurements performed at
RHIC, for meson and baryon pairs. Seen in this light, the validity of
the \pal analysis should be reconsidered if any significant new
effects contributing to such functions are identified. 

The issue of the residual correlations (RC) in femtoscopic
correlations of \bb pairs is mentioned in~\cite{Adams:2005ws}, but the
work explicitly states that it is not addressed and acknowledges this
fact as a weak aspect of the analysis method. In this work we show
that proper treatment of RC is of central importance 
for any \bb measurement, but in particular in the \bab analysis. On
the example of the STAR data we show how the extracted radius and
scattering length change when RC are properly taken into account. We
reanalyze the STAR data with the formalism which includes the RC
contribution. We test, whether the extracted radius is then compatible
with other measurements and model expectations. In the process we make
assumptions on the strong interaction parameters for several \bab
pairs, and show if the extracted values are sensitive to those
assumptions. As a results we put constraints on the \bab strong
interaction parameters, particularly on the imaginary part of the
scattering length, which parametrizes the \bab annihilation process at
low relative momentum.

The paper is organized as follows. In Sec.~\ref{sec:femtoform} we
describe the femtoscopic formalism, including the RC treatment. In
Sec.~\ref{sec:theoryassumptions} we discuss various theoretical
assumptions needed for the reanalysis of the data, and define four
reasonable parameter sets for the theoretical description of the \bab
interaction. In Sec.~\ref{sec:starreana} we examine the STAR data
from~\cite{Adams:2005ws} and show how they should be reanalyzed in the
frame of the formalism taking into account the RC. In
Sec.~\ref{sec:fitex} we apply the formalism to the STAR data and
discuss the results. In Sec.~\ref{sec:summary} we provide the
conclusions and give recommendations for future measurements.  

\section{Femtoscopic formalism}
\label{sec:femtoform}

The femtoscopic correlation function is defined as a ratio of the
conditional probability to observe two particles together, divided by
the product of probabilities to observe each of them
separately. Experimentally it is measured by dividing the distribution
of relative momentum of pairs of particles detected in the same
collision (event) by an equivalent distribution for pairs where each
particle is taken from a different collision. This is the procedure
used by STAR, details are given in~\cite{Adams:2005ws}. The femtoscopy
technique focuses on the mutual two-particle interaction. It can come from
wave-function (anti-)symmetrization for pairs of identical particles,
the measurement in this case is sometimes referred to as ``HBT
correlations''. Another source is the Final State Interaction (FSI),
that is Coulomb or strong. In this work the Coulomb FSI is only
present for \pap pairs, all others are correlated due to the strong
FSI only. The interaction for \pap system is measured in detail and
well described theoretically, we will use existing calculations for
this system and will not vary any of its parameters in the fits. For
details please see~\cite{Lednicky:2005tb}. For all other pairs the
strong FSI is the only source of femtoscopic correlation. Below we
will describe the formalism for the strong interaction only, as it is
the focus of this work. 

In femtoscopy an assumption is made that the FSI of the pairs of
particles is independent from their production. The two-particle
correlation can then be written as~\cite{Lednicky:1981su}:
\begin{equation}
C(\vec{k^{*}}) = {{\int S(\bf{r^{*}}, \vec{k^{*}})
  |\Psi^{S(+)}_{-k^{*}}(\bf{r^{*}}, \vec{k^{*}})|^{2}} \over {\int
  S(\bf{r^{*}}, \vec{k^{*}})}} 
\label{eq:cfrompsi}
\end{equation}
where $\bf{r^{*}}={\bf x}_{1}-{\bf x}_{2}$ is a relative space-time
separation of the two particles at the moment of their
creation. $\vec{k^{*}}$ is the momentum of the first particle in the
Pair Rest Frame (PRF), so it is half of the pair relative momentum in
this frame. $S$ is the source 
emission function and can be interpreted as a probability to emit a
given particle pair from a given set of emission points with given
momenta. The source of the correlation is the Bethe-Salpeter amplitude
$\Psi^{S(+)}_{-k^{*}}$, which in this case corresponds to the solution of
the quantum scattering problem taken with the inverse time
direction. When particles interact with the strong FSI only it can be 
written as:
\begin{equation}
\Psi^{S(+)}_{-k^{*}}({\bf r^{*}}, \vec{k^{*}}) = e^{i\vec{k^{*}}
  \vec{r^{*}}} + f^{S}(k^{*}){{e^{i{k^{*}}{r^{*}}}} \over {r^{*}}}
\label{eq:psidef}
\end{equation}
where $f^{S}$ is the S-wave strong interaction amplitude. In the
effective range approximation it can be expressed as:
\begin{equation}
f^{S}(k^{*}) = \left ({{1} \over {f^{S}_{0}}} + {{1} \over {2}} {d_{0}^{S}
    {k^{*}}^{2}} - ik^{*}\right)^{-1} 
\label{eq:fdef}
\end{equation}
where $f_{0}^{S}$ is the scattering length and $d_{0}^{S}$ is the
effective radius of the strong interaction. These are the essential
parameters of the strong interaction, which can be extracted from the 
fit to the experimental correlation function. Both are complex
numbers; the imaginary part of $f_{0}$ is especially interesting as 
it corresponds to the annihilation process. In the relative momentum
range where the effective range approximation is valid they are also
directly related to the interaction cross-section: $\sigma=4\pi
|f^{S}|^{2}$. Therefore their knowledge is of fundamental importance. 

For one-dimensional correlation function the source function $S$
has one parameter. Usually a spherically symmetric source in PRF with
size $r_{0}$ is taken: 
\begin{equation}
S(\vec{r^{*}}) \approx \exp \left( -{{{r^{*}}^{2}} \over {4 r_{0}^{2}}} \right )
\label{eq:Sdef}
\end{equation}
which gives the final form of the analytical correlation function
depending on the strong FSI only~\cite{Lednicky:1981su,Adams:2005ws}:
\begin{eqnarray}
C(\vec{k^{*}}) &=& 1+\sum_{S} \rho_{S} \left [ {{1} \over {2}} \left |
    {{f^{S}(k^{*})} \over {r_{0}}} \right |^{2} \left (  1-
  {{d_{0}^{S}} \over {2\sqrt{\pi}r_{0}}} \right ) + \right.
\nonumber \\
& & \left.  {{2\Re f^{S}(\vec{k^{*}})} \over
    {\sqrt{\pi}r_{0}}}F_{1}(Qr_{0}) -{{\Im  f^{S}(\vec{k^{*}})} \over {r_{0}}}F_{2}(Qr_{0})
\right ] ,
\label{eq:cstrana}
\end{eqnarray}
where $Q=2k^{*}$, $F_{1}(z) = \int_{0}^{z} dx e^{x^{2} - z^{2}}/z$ and $F_{2} =
(1-e^{-z^{2}})/z$. Summation is done over possible pair spin
orientations, with $\rho_{S}$ the corresponding pair spin
fractions. Since the data considered in this work is always for
unpolarized pairs, the spin dependence of the correlation will be
neglected. In this formula the dependence of the correlation 
function on the real and imaginary part of the scattering length $f_{0}$
is expressed directly. For pairs where only the strong FSI
contributes to the correlation, such as $p\Lambda$ and
$p\bar{\Lambda}$, this formula can be fitted directly to extract the
source size $r_{0}$ as well as the scattering length and effective
radius. In realistic scenarios it is rarely possible to independently
determine all parameters. In particular in case of the STAR data
discussed here the $d_{0}$ was fixed at zero and only the remaining
three were fitted. 

\subsection{Residual correlations}
\label{sec:rescor}

In experiments conducted at colliders such as STAR experiment at RHIC,
all particles propagate to the detector radially from the interaction
point located in the center of the detector. A baryon coming from a
weak decay often travels in a direction very similar to the parent
baryon. The particle's trajectory does not point 
precisely to the interaction point, but this difference (called the
Distance of the Closest Approach or DCA) is often comparable to the
spatial resolution of the experiment. As a result significant number
of particles identified as protons in STAR are not primary and come
from the decay of heavier baryons. The same mechanism applies to
$\Lambda$ baryons. In particular protons can come from a decay of
$\Lambda$ and $\Sigma^{+}$ baryons, while $\Lambda$ baryons can come
from decays of $\Sigma^{0}$ or $\Xi^{0}$. 
%
%
\begin{table}[tb]
\caption{List of possible parent pairs for the $p\Lambda$ (and
  $p\bar{\Lambda}$) system, with their relative contribution to the
  STAR sample~\cite{Adams:2005ws} and the decay momenta values.}
\begin{tabular}{|l|c|c|}
\hline
Pair & Fraction & Decay momenta (MeV/$c$) \\
\hline
$p\Lambda$ & 15\% & 0  \\
$\Lambda\Lambda$ & 10\% & 101  \\
$\Sigma^{+}\Lambda$ & 3\% & 189  \\
$p\Sigma^{0}$ & 11\% & 74  \\
$\Lambda\Sigma^{0}$ & 7\% & 101, 74  \\
$\Sigma^{+}\Sigma^{0}$ & 2\% & 189, 74  \\
$p\Xi^{0}$ & 9\% & 135  \\
$\Lambda\Xi^{0}$ & 5\% & 101, 135  \\
$\Sigma^{+}\Xi^{0}$ & 2\% & 189, 135  \\
$pp$ & 7\% & 101  \\
\hline
\end{tabular}
\label{tab:pairlist}
\end{table}
STAR experiment has applied the DCA cut to reduce the number of
such secondaries and has estimated its effectiveness based on the
Monte-Carlo simulation of the detector response. The fraction of
true primary pairs, as well as a fraction of all other parent particle
pair combinations is taken from~\cite{Adams:2005ws} and given in
Tab.~\ref{tab:pairlist}. In addition to the effect mentioned above it
is also possible that a primary proton is randomly associated to a
pion and reconstructed as a fake $\Lambda$ baryon. For that reason a
pair of two protons also appears in Tab.~\ref{tab:pairlist}.

The strong FSI affects the behavior of the two particles in the pair
just after their production, on a time scale of fm/$c$. For particles
coming from a weak decay, which occurs on timescales of $10^{-10}$~s,
the FSI applies to the parent pair, not the daughter. However it is the
daughters that are measured in the detector. For such a scenario
Eq.~\eqref{eq:cfrompsi} cannot be used directly. In this case $\Psi$
must be taken for the parent pair and calculated for $k^{*}$ and
$r^{*}$ between the parent particles. Then one or both of the parent
particles must decay and a new $k^{*}$ must be calculated for the
daughter pair. This one is measured in the detector, the correlation is
measured as a function of this relative momentum. Such scenario is
called ``residual correlations''
(RC)~\cite{Wang:1999bf,Wang:1999xq}. Obviously the random nature of
the weak  
decay will dilute the original correlation. However if the decay
momentum is comparable to the width of the correlation effect in
relative momentum, some correlation might be preserved for the
daughter particles. The RC are important if three conditions
are met simultaneously: a) the original correlation for parent
particles is large, b) the fraction of daughter pairs coming from a
particular parent pair is significant and c) the decay momentum (or
momenta) are comparable to the expected correlation width in
$k^{*}$. For $p\bar{\Lambda}$ pairs all three conditions are met. The
strong FSI for baryon-antibaryon pairs is dominated by annihilation,
which appears in Eq.~\eqref{eq:cstrana} as $\Im f^{S}$. It causes a
negative correlation (anticorrelation), which is wide in $k^{*}$, even
on the order of $300$~MeV/$c$. Decay momenta for all residual pairs
listed in Tab.~\ref{tab:pairlist} are of that order or
smaller. Comparing contributions to the sample from all pairs one can
see that all listed are of the same order as primary $p\bar{\Lambda}$
pairs which constitute only 15\% of the sample. As for the strength of
the correlation, it is in principle unknown for all pairs, except
$p\bar{p}$. Estimating its strength is one of the goals of this
work. However it is often assumed that at least the annihilation
cross-section for \bab pairs is very similar for all pairs, comparable
to \pap~\cite{Bleicher:1999xi}. In that case it is certainly strong
enough to induce RC and contributions from all pairs listed in
Tab.~\ref{tab:pairlist} must be considered in the analysis of the \pal
correlations. 

The RC can be calculated for any combination of parent and daughter
pairs. The correlation is expressed as a function of the relative
momentum of the daughter pair, in our case it is
$k^{*}_{p\Lambda}$. However Eq.~\eqref{eq:cfrompsi} is then used for
the parent pair (let's call it $XY$), and gives the correlation as a
function of $k^{*}_{XY}$. Baryon $X$ is a proton or decays into a
proton and baryon $Y$ is a $\Lambda$ or decays into a $\Lambda$. The
daughter momenta will differ from the 
parents' by the decay momentum, listed in Tab.~\ref{tab:pairlist}. The
direction of the decay momentum is random in the parents' rest frame,
and it is independent from the direction of $k^{*}$ of the
pair. Therefore $k^{*}_{p\Lambda}$ will differ, in a random way for
each pair, from $k^{*}_{XY}$. The difference is limited by the
value of the decay momentum and is a non-trivial consequence of the
decay kinematics. 

\begin{figure}[tb]
\begin{center}
\includegraphics[angle=0,width=0.45 \textwidth]{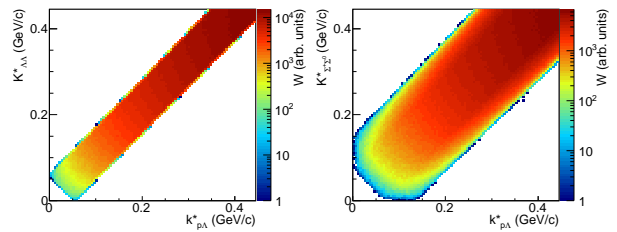}
\end{center}
\vspace{-6.5mm}
\caption{The unnormalized transformation matrix W for $\Lambda\Lambda$
  (left) and $\Sigma^{+}\Sigma^{0}$ (right) pairs decaying into
  $p\Lambda$ pairs, as a function of relative momentum of both pair types.
\label{fig:wmatexamples}}
\end{figure}

One can determine what is the probability that a parent particle pair
with a given $k^{*}_{XY}$ will decay into a daughter pair with a given
$k^{*}_{p\Lambda}$. Let's call such probability distribution
$W(k^{*}_{XY}, k^{*}_{p\Lambda})$. In this work we have calculated it
for all pairs listed in Tab.~\ref{tab:pairlist}. We have used the
Therminator model~\cite{Kisiel:2005hn, Chojnacki:2011hb}, with
parameters describing central Au+Au collisions at $\sqrt{s_{NN}} =
200$~GeV. All the pairs of type $XY$ in a given event were found and
their relative momentum $k^{*}_{XY}$ was calculated. Than both baryons
$X$ and $Y$ were allowed to decay and $k^{*}_{p\Lambda}$ was
calculated for the daughters. The pair was then inserted in a
two-dimensional histogram. As a result an unnormalized probability
distribution $W$ was obtained for each pair
type. Fig.~\ref{fig:wmatexamples} shows two examples of this function,
one for a pair where only one particle decays, the other for a pair
where both particles decay. In the first case the function has a
characteristic rectangular shape at low relative
momentum~\cite{Wang:1999bf,Wang:1999xq}. It touches both axes 
at the value roughly equal to half of the decay momentum. The vertical
width of the function is roughly equal to the decay momentum, as
discussed above. In the second case the shape at low momentum is not
as sharp, and the width is equal to the sum of decay momenta. $W$
depends only on decay kinematics, so it is the same for \bb and the
corresponding \bab pair.

\begin{figure}[tb]
\begin{center}
\includegraphics[angle=0,width=0.45 \textwidth]{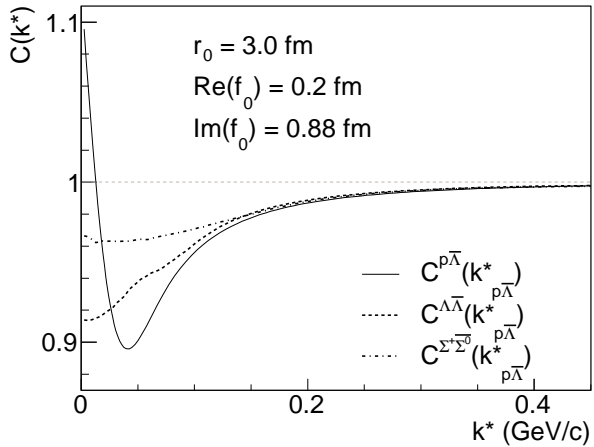}
\end{center}
\vspace{-6.5mm}
\caption{Theoretical correlation function for a given source size for
  $p\bar{\Lambda}$ and two examples of residual correlation functions for
  $\Lambda\bar{\Lambda}$ and $\Sigma^{+}\bar{\Sigma^{0}}$ pairs.
\label{fig:crcexamples}}
\end{figure}

Having defined $W$ one can write the formula for the RC for any type
of the parent pair $X\bar{Y}$, contributing to the \pal correlation 
function:
\begin{equation}
C^{X\bar{Y} \rightarrow p\bar{\Lambda}}(k^{*}_{p\bar{\Lambda}}) = {{\int
    C^{X\bar{Y}}(k^{*}_{X\bar{Y}})
    W(k^{*}_{X\bar{Y}},k^{*}_{p\bar{\Lambda}}) d k^{*}_{X\bar{Y}} } \over
  {\int W(k^{*}_{X\bar{Y}},k^{*}_{p\bar{\Lambda}})d k^{*}_{X\bar{Y}} }} .
\label{eq:rcdef}
\end{equation}
Examples of correlation functions transformed in this way are shown in
Fig.~\ref{fig:crcexamples}. The \pal function for a given source size,
calculated according to Eq.~\eqref{eq:cstrana} is given for
comparison. It has positive correlation at very low $k^{*}$ coming
from the positive real part of the scattering length $f_{0}$ and a
wide anticorrelation coming from the positive imaginary part of
$f_{0}$. This anticorrelation is wide, extending beyond $0.4$~GeV/$c$,
so its width is larger than any combination of decay momenta given in
Tab.~\ref{tab:pairlist}. The residual correlations are calculated for
the same source size and radius parameters, so in their respective
$k^{*}$ variables they look identical to
$C^{p\bar{\Lambda}}(k^{*}_{p\bar{\Lambda}})$. After the transformation
given by Eq.~\eqref{eq:rcdef} the correlation is diluted at low 
$k^{*}$. However at higher values the shape of the function changes
very little and is almost the same for the parent and residual
correlation. The spike at $k^{*}=0$ is transformed with the matrix in
Fig.~\ref{fig:wmatexamples} to a slight bump in
$k^{*}_{p\bar{\Lambda}}$ where $W$ 
touches the $x$ axis, that is around 50 
MeV/$c$, half of the decay momentum of $\Lambda$ into proton. The same
function is diluted twice as strong at low 
$k^{*}$ for the pair where both particles decay
($\Sigma^{+}\bar{\Sigma^{0}}$). This difference persists up to
around 100 MeV/$c$, above this value both functions are similar to
each other and to the original correlation. 

In terms of the physics
picture the contribution of the \pap correlation to the \pal one is
not RC, instead it comes from fake association of primary proton to
a $\Lambda$ particle. However the formalism to deal with such
situation is exactly the same as in the case of RC and Eq.~\eqref{eq:rcdef}
can be used. The difference is that the \pap correlation function has
a Coulomb FSI component in addition to the strong FSI, which must be
taken into account when calculating $C^{p\bar{p}}$. The $W$
matrix for $p\Lambda$ to $pp$ pair transformation can be used. 

Once each of the RC components is determined, the complete correlation
function for the \pal system can be written:
\begin{eqnarray}
C(k^{*}_{p\bar{\Lambda}}) &=& 1 +
\lambda_{p\Lambda}\left(C^{p\bar{\Lambda}}(k^{*}_{p\bar{\Lambda}})-1 
\right) \nonumber \\
& &+ \sum_{X\bar{Y}} \lambda_{XY}\left(C^{X\bar{Y}}(k^{*}_{p\bar{\Lambda}})-1
\right)
\label{eq:cfallrc}
\end{eqnarray}
where the $\lambda$ values are equivalent to the pair fractions given
in Tab.~\ref{tab:pairlist}. It is an additional factor that decreases
the correlation for the RC, however for some pairs it is almost as
large as $\lambda$ for true \pal pairs. Eq.~\eqref{eq:cfallrc} is the
final formula that can be fitted directly to experimental
data. In principle each $C^{X\bar{Y}}$ depends on four  
independent parameters: the source size $r_{0}$, real and imaginary
part of $f_{0}$ and the value of $d_{0}$, giving 37 independent
parameters ($f_{0}$ and $d_{0}$ for the \pap pair is known). Some
assumptions are obviously 
needed to reduce this number, we will propose several options in 
Sec.~\ref{sec:theoryassumptions}.

\section{Theoretical scenarios}
\label{sec:theoryassumptions}

Following the procedure employed by STAR in~\cite{Adams:2005ws} we
put the effective range $d_{0} = 0$~fm for all calculations. Radius
for the  various systems in central Au+Au collisions at RHIC energies
is expected to follow hydrodynamic predictions, which give $r_{0} \sim
1/\sqrt{\left <m_{T}\right>}$, where $m_{T}$ is the transverse mass of
the pair. For baryons $m_{T}$ is large, and the decrease is not
expected to be steep (see Fig.~5 in~\cite{Adams:2005ws} for
illustration). $\left < m_{T} \right>$ for a given pair 
depends on the momentum spectra of particles taken for this analysis,
which is not specified in~\cite{Adams:2005ws}. We expect that $\left
  <m_{T} \right>$ for the pairs considered here will be within 20\% of
each other, giving little variation of the scaling factor. Therefore
we make a  simplifying assumption that system size $r_{0}$ for each
pair is the same. 

With these assumptions 18 components of $f_{0}$ remain for the nine
pairs. Little theoretical guidance is given for those values. An
approach adopted in~\cite{Bleicher:1999xi} equates all annihilation
cross-sections for the \bab pairs and assumes they are equal to the
one for $p\bar{p}$. In~\cite{Batty:1989gg} the value of $\Im{f_{0}} = 0.88 \pm
0.09$~fm is given. This value is used to calculate the \pap
correlation functions. No such assumption is made for $\Re{f_{0}}$,
which can vary significantly between various \bab pairs.
Therefore, we make two assumptions. $\Im{f_0}$ is assumed to be the
same for all \bab pairs, but it is not fixed to the \pap value - it is
treated as free in the fit. Similarly $\Re{f_0}$ is also assumed to be
the same for all pairs and is free in the fit.

In~\cite{Bleicher:1999xi} an alternative scenario for annihilation
cross-sections is given. Namely that they are the same as in $p\bar{p}$, but
at the same $\sqrt{s}$ of the pair, not relative momentum. In UrQMD
these assumptions differ little, the majority of hadronic
rescatterings happen at large relative momentum, where the
difference between cross-sections scaled with $k^{*}$ and $\sqrt{s}$
is small. However in the case 
of femtoscopic correlations, which by definition are concentrated at
low relative momentum, the two scenarios differ strongly. For
example a $\Sigma^{+}\Xi^{0}$ pair at $k^{*}=10$~MeV/$c$ taken at the
same $\sqrt{s}$ corresponds to a $pp$ pair at
$k^{*}=831.6$~MeV/$c$. This assumption would then significantly
decrease the correlation for higher-mass pairs, including $p\bar{\Lambda}$. We 
test whether it is consistent with the data.
%
%

The next scenario comes from the fact, that we have just shown that
both \pap and $\Lambda\bar{\Lambda}$ RC will contribute to the \pal
correlation function. One can then ask if it is possible that the
observed correlation is explained by annihilation of
particle-antiparticle pairs only, not all \bab pairs. In such
scenario the imaginary part of the scattering length should be put to
zero for all pairs except the ones in which the two particles have exactly
opposite quark content. 

The last scenario is the repetition of the STAR procedure, where no RC
is included and correlation is present for \pal pairs only. 

In each of these four scenarios, all $C^{X\bar{Y}}$
functions can be calculated from Eq.~\eqref{eq:cstrana}. The last
function remaining to be calculated is then $C^{p\bar{p}}$. A dedicated
procedure is used. First the relative momenta distributions are taken from
Therminator, from collisions simulated with parameters corresponding
to central Au+Au collision at $\sqrt{s_{NN}}=200$~GeV. Then a source
size is assumed, equal to the one used for all other pairs. This
allows the generation of $r^{*}$ for each pair, according to the
probability distribution from Eq.~\eqref{eq:Sdef}. This gives 
pairs, each with its $k^{*}$ and $r^{*}$, which enables the
calculation of $\Psi$. It is performed with a dedicated code from
Lednicky~\cite{Lednicky:2005tb}, where the known \pap interaction
parameters are used. The resulting correlation function is then
calculated according to Eq.~\eqref{eq:cfrompsi}, corresponding to the
value of the source size $r_{0}$. As mentioned earlier, the $W$
functions are also calculated from Therminator, for all parent pair
types. 

\section{Analysis of STAR data}
\label{sec:starreana}

The STAR data on \pal correlation function~\cite{Adams:2005ws} has
been corrected for several effects, most of them experimental in
nature. Two of those corrections must be reexamined for this
analysis. The correlation was normalized ``above 0.35
GeV/$c$''~\cite{Adams:2005ws}. As 
can be see in Fig.~\ref{fig:crcexamples} the femtoscopic correlation
is small but non-negligible in this region. However the upper range
for the normalization is not given. The number of pairs increases with
$k^{*}$, so if the upper normalization range is large, pairs with
negligible correlation will dominate the normalization factor. We will
assume this is the case, which means that the experimental correlation
is properly normalized.  

The data was also corrected for ``purity'', that is the fraction of
true \pal pairs, given in Tab.~\ref{tab:pairlist}. The procedure used
by STAR is correct only if all other pairs are not correlated. In
Eq.~\eqref{eq:cfallrc} it would correspond to the scenario where all
$C^{X\bar{Y}}$ are at 1.0 in the full $k^{*}$ range. We have
just shown that this assumption is explicitly violated by the RC
effect, which is expected to be significant for \pal
correlations measured by STAR. 
Therefore
the experimental correlation function analyzed later will be
``uncorrected'' for purity, with the purity factor equal to 0.15,
taken from Tab.~\ref{tab:pairlist}, so that a fit according to
Eq.~\eqref{eq:cfallrc} can be properly applied.   

The fitting range was set to $0.45$~GeV/$c$, the maximum range for
which experimental data is available.

\section{Fitting the experimental correlation function}
\label{sec:fitex}

\begin{figure}[tb]
\begin{center}
\includegraphics[angle=0,width=0.45 \textwidth]{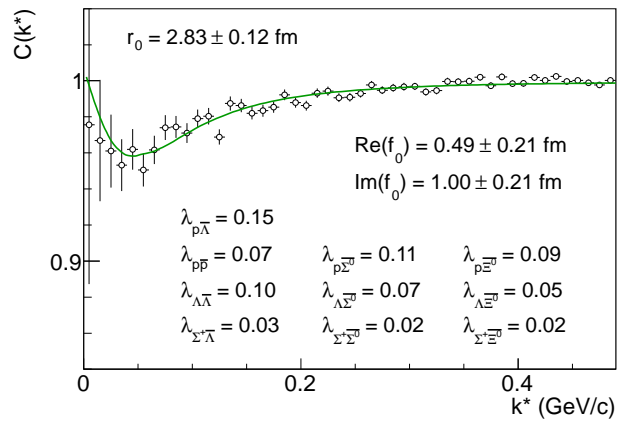}
\end{center}
\vspace{-6.5mm}
\caption{Fit to the STAR $p\bar{\Lambda}$ correlation function with
  Eq.~\eqref{eq:cfallrc}, all residual correlation components included.
\label{fig:plafullrc}}
\end{figure}

Formula~\eqref{eq:cfallrc} is fitted to the STAR experimental data,
with the theoretical assumptions mentioned above. Standard $\chi^2$
minimization procedure is used. The result of the fit is shown
in Fig.~\ref{fig:plafullrc}. It gives the value of the source size
$r_{0} = 2.83 \pm 0.12$~fm, and the scattering length $f_{0} = 0.49
\pm 0.21 + i (1.00 \pm 0.21)$~fm. The value of $r_{0}$ is
significantly larger than given in~\cite{Adams:2005ws}, indicating
that the RC play a critical role in the extraction of physical
quantities. The value extracted here is in good agreement with the
values obtained for the \pla system. This consistency is naturally
expected in practically all realistic models of heavy-ion collisions,
while the previous STAR result was violating this consistency without
providing any viable explanation. It is also consistent with
expectation from hydrodynamical models, which are in good agreement
with all other femtoscopic measurements at RHIC. Taking all those
arguments into account we claim that the result presented in this work
is the correct one, and that the result for \pal
from~\cite{Adams:2005ws} should be considered obsolete.  

The extracted imaginary part of the scattering length is significant
and in agreement with the value given for the \pap system. This means
that the assumption that the annihilation process for any \bab system
is similar to that process for \pap, taken at the same relative
momentum is consistent with data. 

\begin{figure}[tb]
\begin{center}
\includegraphics[angle=0,width=0.45 \textwidth]{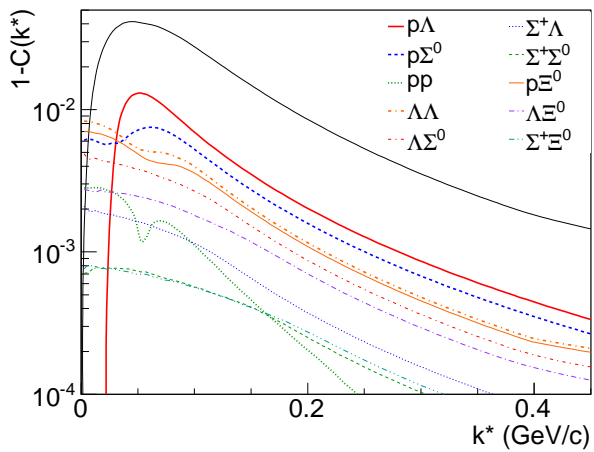}
\end{center}
\vspace{-6.5mm}
\caption{Comparison of all residual correlation components for the
  $p\bar{\Lambda}$ correlation function (thin black line). For better
  illustration the inverse of the correlation effect is plotted.
\label{fig:rccomponents}}
\end{figure}

In Fig.~\ref{fig:rccomponents} all the residual correlation components
of the fit are shown. The absolute value of the correlation effect
$1-C$ is plotted, the logarithmic scale is needed to distinguish the
small contributions. No single component is dominating the function,
all 10 components are needed to describe the correlation. The largest
ones are, as expected, the ones which have large pair fractions and small
decay momenta, that is $p\bar{\Lambda}$, $p\bar{\Sigma^{0}}$,
$\Lambda\bar{\Lambda}$ and $p\bar{\Xi^{0}}$. The systems where both
particles decay and the systems where the fraction is small contribute
less. All the RC contributions are relevant through the whole $k^{*}$
range.   

\begin{figure}[tb]
\begin{center}
\includegraphics[angle=0,width=0.45 \textwidth]{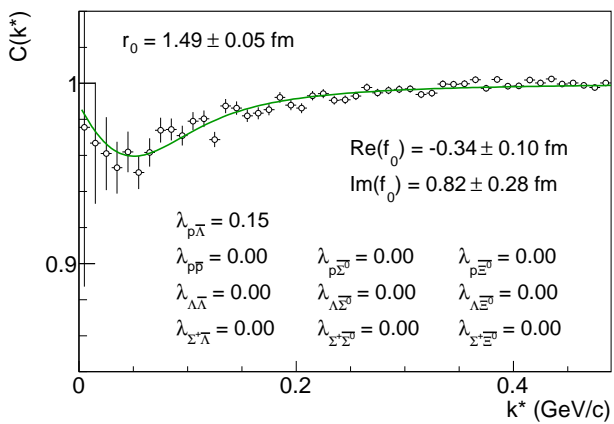}
\end{center}
\vspace{-6.5mm}
\caption{Fit to the STAR $p\bar{\Lambda}$ correlation function with
  Eq.~\eqref{eq:cfallrc}, no residual correlation components included.
\label{fig:planorc}}
\end{figure}

In order to validate the procedure and the new important result,
several scenarios, described in Sec.~\ref{sec:theoryassumptions}, have
been tested. In Fig.~\ref{fig:planorc} the fit 
was performed, where no residual correlations were included. This is
equivalent to the STAR procedure. The result from~\cite{Adams:2005ws}
is reproduced, the resulting radius is small. $\Re{f_0}$ changes sign
with respect to the default case, but interestingly $\Im{f_0}$ is 
consistent with the full RC fit. 

The next scenario assumes annihilation for particle-antiparticle pairs
only. By testing it we check if the annihilation is really necessary
for all \bab pairs, or is enough if it happens only with baryons
having exactly the opposite quark content. A fit is performed, where
only \pap and $\Lambda\bar{\Lambda}$  RC is included, while for all
other \bab pairs (including $p\bar{\Lambda}$) there is no
correlation. Result similar to the previous test is obtained - the
radius is 
$1.5 \pm 0.1$~fm. Both scenarios are therefore unlikely. In other
words the analysis shows that the annihilation happens between all
\bab pairs, not just the ones with exactly opposite quark content and
that this effect must be taken into account, via the RC formalism in
any analysis of \bab femtoscopic correlations.
 


\begin{figure}[tb]
\begin{center}
\includegraphics[angle=0,width=0.45 \textwidth]{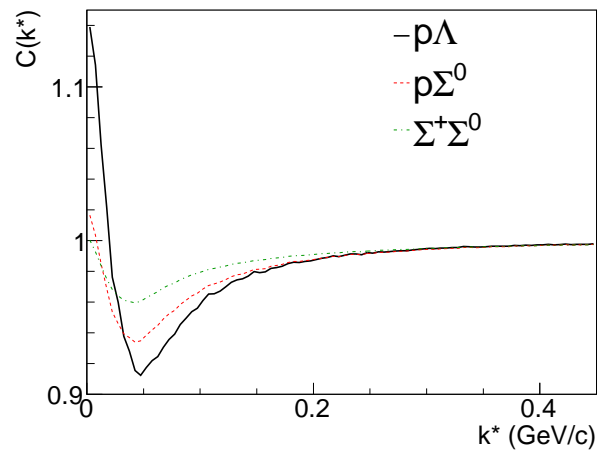}
\end{center}
\vspace{-6.5mm}
\caption{Calculation of correlation function, with $f^{S}$ taken the
  same as the $f^{S}$ for \pal pair at the corresponding $\sqrt{s}$.
\label{fig:ffromqcf}}
\end{figure}

In the last scenario, following the idea from~\cite{Bleicher:1999xi}
it was proposed that the annihilation cross-section for \bab pairs is
the same for all pairs, but taken at the same $\sqrt{s}$ instead of the
relative momentum. In femtoscopy such scaling would be reflected in
Eq.~\eqref{eq:psidef} by taking $f^{S}$ at a different  $k^{*}$. In this
work we treat the imaginary and real parts of $f_{0}$ for the \pal 
system as fit parameters and scale the $f^{S}$ for all other pairs, by
taking the same $f_{0}$ parameters, but calculating $f^{S}$ at:
\begin{equation}
k^{*} = \left ( {{s^{2} +m_{p}^{4} + m_{\Lambda}^{4} - 2 s m_{p}^{2} -
      2 s m_{\Lambda}^{2} - 2m_{p}^{2}m_{\Lambda}^{2}} \over {4 s}}
\right )^{1/2}
\label{eq:kscale}
\end{equation}
according to Eq.~\eqref{eq:fdef}. $s$ is the square of the total energy
in PRF for the pair $X\bar{Y}$. $f^{S}$ is a function rapidly
decreasing with $k^{*}$. By taking $s$ for the baryon pair, where one
or both baryons have a mass higher than the proton or the $\Lambda$,
one gets from Eq.~\eqref{eq:kscale} $k^{*}$ higher than for the original
pair, so $f^{S}$ will be smaller. In Fig.~\ref{fig:ffromqcf} the
result of such calculation is shown for a pair with smallest and
largest mass difference to the \pal pair. The strength of the
correlation is visibly decreased. However the shape is only slightly
affected. In fact the functions can be described by
Eq.~\eqref{eq:cstrana}, with altered values of $f_{0}$. The $\Re{f_{0}}$
is scaled to approximately 20\% of the original value, while
$\Im{f_{0}}$ is scaled to 60\% (32\%) of the original value for the
pair with smallest (largest) mass difference, that is $p\Sigma^{0}$
($\Sigma^{+}\Xi^{0}$). These scaling factors provide the needed
constraints on the fit parameters, and the fit can be performed as in
the previous cases.

\begin{figure}[tb]
\begin{center}
\includegraphics[angle=0,width=0.45 \textwidth]{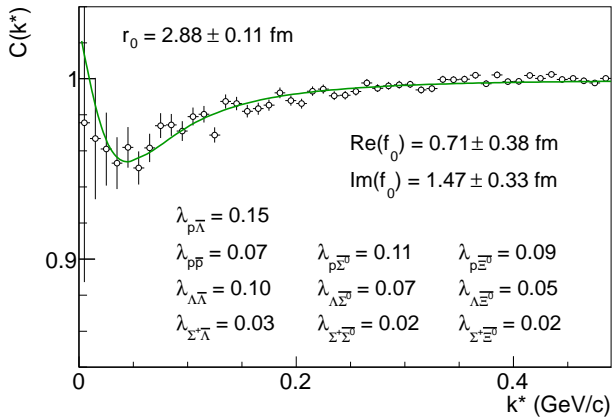}
\end{center}
\vspace{-6.5mm}
\caption{Fit the the STAR $p\bar{\Lambda}$ correlation function with
  Eq.~\eqref{eq:cfallrc}, all residual correlations included, strength
  of the interaction scaled according to the $\sqrt{s}$ of the pair
  (see text for details). 
\label{fig:plarcsscale}}
\end{figure}

Fig.~\ref{fig:plarcsscale} shows the result of the fit, with the scaling
of $f^{S}$ with $\sqrt{s}$ of the pair. The resulting source size is
comparable to the default case. $\Im{f_{0}}$ is significantly larger
than for the default fit and larger than the measured \pap
value. While this scenario is not ruled out by the data, it is
internally inconsistent. It would mean that moving from \pap to
heavier pairs, the cross-section first increases sharply and then
decreases for heavier pairs. If one takes the \pap $f_{0}$ as the
starting point, instead of \pal (which would be a more literate
implementation of the scenario proposed in~\cite{Bleicher:1999xi}),
then $f_{0}$ cannot be a free parameter. A fit gives $r_{0}=2.23 \pm
0.09$~fm which is lower than the expected value. Such scenario cannot
be ruled out, but is less likely, due to the disagreement of this
value with $r_{0}$ for \pla pairs.  

\subsection{Systematic uncertainty discussion}
\label{sec:systunc}

All the values given above were obtained with certain assumptions,
spelled above, both related to the STAR data treatment as well as
the methodology itself and the unknown strong interaction
parameters. By varying those assumptions in a reasonable range one can
estimate the systematic uncertainty on the extracted parameters coming
from the application of the RC method and the assumptions made. 

Restricting the fitting range to $0.35$~GeV/$c$ (beginning of the
normalization range) gives 5\% variation in radius, while $\Im{f_{0}}$
decreases to $0.6 \pm 0.2$~fm and $\Re{f_{0}}$ is positive but
consistent with zero. Performing the fit separately for \pal and \apla
pairs gives $r_{0}$ statistically consistent with the default
fit. $\Im{f_{0}}$ varies by up to 20\%, and $\Re{f_{0}}$ by up to
50\%. That is expected - $\Re{f_{0}}$ affects the function mostly at
low $k^{*}$, where data is less precise, while $\Im{f_{0}}$ produces
the wide anticorrelation which is better constrained by the data. With
the statistical power of the STAR data we were 
unable to test the influence of the $d_{0}$ parameter variation, or
independent variation of $f_{0}$ parameters for heavier \bab pair
types. In conclusion the source size $r_{0}$ is well constrained and
comparable to $r_{0}$ measured by STAR~\cite{Adams:2005ws} for \pla
and \apal within the statistical and systematic uncertainty of this
work. $\Im{f_{0}}$ is determined to be finite and positive, consistent
with the hypothesis that its value for all  \bab pairs considered is
similar to the value for $p\bar{p}$. The systematic uncertainty of the
method is at least 20\%. $\Re{f_{0}}$ is consistent with being finite and
positive, although the systematic uncertainty of the method is at
least 50\%. There is also no theoretical expectation that $\Re{f_{0}}$
is similar for different \bab pairs, so this measurement can be
interpreted as ``average effective'' $\Re{f_{0}}$ for the considered
\bab pairs. 

Certain other systematic uncertainties depend on the detail of the
experimental treatment. These include, among others, the variation of
the normalization range, variation of the pair fractions and the
variation of the DCA cuts. Their estimation is beyond the scope of
this work, as it requires direct access to experimental raw data and
procedures.  

\section{Summary}
\label{sec:summary}

We have presented the theoretical formalism for dealing with residual
correlations in baryon-antibaryon femtoscopic correlations. We have
shown that for realistic scenario of heavy-ion collision at
$\sqrt{s_{NN}}=200$~GeV such correlations are critical for the
correct interpretation of data. The formalism has been applied to \pal
and \apla femtoscopic correlations measured by
STAR~\cite{Adams:2005ws}. New estimates for system size $r_{0}$ as
well as real and imaginary parts of the scattering length $f_{0}$ have
been obtained. New system size is consistent with results for \pla and
\apal pairs and model expectations. Therefore the puzzle of
unexpectedly small \pal system size reported by STAR
in~\cite{Adams:2005ws} is solved. In addition new, more robust
estimates for $f_{0}$ parameter is obtained, not only for the \pal system,
but also for a number of heavier \bab pairs. A scenario where all \bab
pairs have similar annihilation cross-section (expressed as a function
of pair relative momentum) is judged to be most likely, as it gives
the expected source size and is internally consistent. Other scenarios
have been explored, but were judged to be less likely. 

With the new methodology it is possible to measure strong interaction
potential for a number of \bab pair types, including $\Lambda$ and
$\Xi$ baryons. More precise data, differential in centrality and pair
momentum and obtained for other pair types (e.g. $p\Xi^{0}$,
$\Lambda\Lambda$, $\Lambda\Xi^{0}$) would help constrain this
interesting, unknown quantities. In particular high statistics
runs of Au+Au collisions at RHIC, as well as Pb--Pb collisions at the
LHC promise better quality data and give hope for more precise
measurement in the near future. 

\bibliography{citations}

\end{document}